\documentstyle[epsf,eqsecnum,floats,preprint,aps]{revtex}

\input epsf
\tighten
\newcommand{\PSbox}[3]{\mbox{\rule{0in}{#3}\includegraphics{#1}\hspace{#2}}}

\def\bsigma{\mbox{\boldmath $\sigma$}}

\begin{document}

\rightline{CU-TP-1012}
\rightline{hep-th/0105211}
\vskip 1cm

\begin{center}
\ \\
\large{{\bf Scattering of massless and massive monopoles in an SU($N$)
theory }} 
\ \\
\ \\
\ \\
\normalsize{Xingang Chen\footnote{\tt xgchen@phys.columbia.edu} and Erick J. Weinberg\footnote{\tt ejw@phys.columbia.edu}}
\ \\
\ \\
\small{\em Department of Physics \\

Columbia University \\
New York, NY 10027}

\end{center}

\begin{abstract}
We use the moduli space approximation to study the time evolution of
magnetically charged configurations in a theory with an SU($N+2$)
gauge symmetry spontaneously broken to
U(1)$\times$SU($N$)$\times$U(1).  We focus on configurations
containing two massive and $N-1$ massless monopoles.  The latter do
not appear as distinct objects, but instead coalesce into a cloud of
non-Abelian field.  We find that at large times the cloud and the
massless particles are decoupled, with separately conserved energies.
The interaction between them occurs through a scattering process in
which the cloud, acting very much like a thin shell, contracts and
eventually bounces off the cores of the massive monopoles.  The
strength of the interaction, as measured, e.g., by the amount of
energy transfer, tends to be greatest if the shell is small at the
time that it overlaps the massive cores.  We also discuss the
corresponding behavior for the case of the SU(3) multimonopole
solutions studied by Dancer.   
\end{abstract}

\setcounter{page}{0}
\thispagestyle{empty}
\maketitle

\eject

\vfill

\baselineskip=18pt

\section{introduction}

There has long been an interest in the magnetic monopoles that arise
as classical solutions in certain spontaneously broken gauge theories.
When the theory is quantized, these give rise to magnetically charged
particles that can be regarded as the counterparts of the electrically
charged elementary quanta of the theory.  In certain supersymmetric
theories, these two classes of particles are believed to be related by
an exact duality symmetry \cite{montonen}.  

When the gauge symmetry is spontaneously broken, the gauge bosons
corresponding to the generators of the unbroken subgroup remain
massless.  If the unbroken subgroup contains a non-Abelian factor,
some of these massless gauge bosons transform nontrivially under
the gauge group and thus carry an ``electric'' charge.  Duality
suggests that these should have massless magnetically-charged
counterparts.  Although these cannot be realized as isolated classical
solutions, evidence for their existence can be found by analyzing
multimonopole solutions.  In this paper we will investigate some of
the properties of these massless monopoles by examining the role that
they play in low energy scattering processes.\footnote{Although the
underlying motivation for this work arises from theories with extended
supersymmetry, this supersymmetry does not come into play at the
leading, classical, order to which we work.}

Recall that an arbitrary magnetically charged
Bogomol'nyi-Prasad-Sommerfield (BPS) \cite{bps} solution can be
naturally understood as being composed of a number of fundamental
monopoles of various types \cite{Gindex}.  If the gauge symmetry $G$
is broken to an Abelian subgroup U(1)$^r$, as happens for generic
choices of the adjoint Higgs vacuum expectation value, there is an
integer-valued topological charge for each U(1) factor.  Associated to
each of these is a fundamental monopole carrying a single unit of that
topological charge.  Each of these fundamental monopoles can be
realized as a classical solution by embedding the unit SU(2) monopole
solution in an appropriate subgroup of $G$.

For the special choices of the Higgs vacuum expectation value that
give a non-Abelian unbroken subgroup, the BPS mass formula implies
that some of the fundamental monopoles should become massless.
Although the corresponding embedding solutions reduce to pure vacuum
solutions in this limit, analysis of multimonopole solutions suggests
that the collective coordinates of the massless fundamental monopoles
survive as degrees of freedom.  Examining static classical solutions
with magnetic charges corresponding to a sum of massive and massless
monopoles, one finds that the massive monopoles are surrounded by one
or more ``clouds'' of non-Abelian field \cite{so5}.  The number of
collective coordinates \cite{nonabelindex} needed to describe these clouds
is exactly equal to the number that one would have attributed to the
massless monopoles.\footnote{This counting of collective coordinates
assumes that the total magnetic charge of the configuration is
Abelian.  Configurations whose total magnetic charges have non-Abelian
components can be regarded as having clouds that have expanded out to
spatial infinity, and have a corresponding reduction in the number of
collective coordinates.  For discussions of some of the pathologies
that arise in the presence of non-Abelian magnetic charges, see
\cite{pathologies}.}

To make these ideas more concrete, consider the case of an SU($N+2$)
gauge theory with an adjoint Higgs field whose asymptotic value (in
some fixed direction) can be brought into the form
\begin{equation}
    \Phi = {\rm diag}\, (t_1, t_2, \dots, t_{N+2})
\label{asymHiggs}
\end{equation} 
with $t_1 \ge t_2 \ge \dots t_{N+2}$.  The asymptotic magnetic field
in the same direction can be written as $F_{ij} =
\epsilon_{ijk}r_kQ_M/r^3$, where the magnetic charge $Q_M$ is of
the form
\begin{equation}
   Q_M = {\rm diag}\,(n_1, n_2-n_1, \dots, n_{N+1}-n_N, - n_{N+1})
   \,  \,.
\label{asymQM}
\end{equation} 
The $n_k$ are all integers, and give the number of fundamental
monopoles of the various types.  The mass of the $k$th species of
fundamental is proportional to $t_k -t_{k+1}$.  Hence, if the $t_j$
are all distinct, so that the symmetry is broken maximally, to
U(1)$^{N+1}$, the $N+1$ species of fundamental monopoles are all
massive.  If $M\ge 2$ of the $t_k$ are equal, there is an enlarged
unbroken symmetry, with $M-1$ of the U(1) factors being replaced by an
SU($M$).  

We will focus on solutions containing one of each species of monopole,
with $t_1 \ne t_2=t_3 =\dots = t_{N+1} \ne t_{N+2}$.  The unbroken
group is then U(1)$\times$SU($N$)$\times$U(1), and our solutions
contain two massive and $N-1$ massless monopoles.  This example has
two advantages for us.  First, a number of analytic results about the
static solutions and their moduli space are already known.  Second,
because the two massive monopoles correspond to commuting SU(2)
subgroups of the SU($N+2$), they can have no direct interactions.  Any
interaction between them must be due to the mediation of the massless
monopoles, and thus can give us insight into the properties of the
cloud.

Explicit solutions for an SU($N+2$) theory with the $n_k$ all
equal to unity were obtained in Ref.~\cite{ewyi}.  For either type of
symmetry breaking, these solutions are described by $4(N+1)$ collective
coordinates.  When the fundamental monopoles are all massive, these
have a natural interpretation as the spatial coordinates, ${\bf x}_a$,
and U(1) phases of the $N+1$ constituent monopoles; as long as the
separation between the monopole positions is large compared to the
monopole core sizes, the multimonopole nature of the configuration is
evident in the classical solution.  As the Higgs field approaches the
value corresponding to the non-Abelian breaking, monopoles 1 and
$N+1$, which remain massive, clearly retain their
identity.\footnote{We label the component monopoles by the particular
magnetic charge that they carry; thus monopole $a$ has $n_a=1$ and
corresponds to an SU(2) embedding in the $a$ and $a+1$ rows and
columns of the SU($N+2)$ matrices.}  However, the $N-1$ monopoles that
become massless are replaced by a single cloud surrounding the massive
monopoles; this cloud can be viewed as being formed by the merger of
the cores of the massless monopoles.  Inside this cloud, the
magnetic field is approximately equal to the the Coulomb field
appropriate to the charges of the massive monopoles; it has both
Abelian and non-Abelian components. Outside the cloud, the Coulomb
component of the magnetic field is purely Abelian, with the
non-Abelian contributions falling with a higher power of distance.

The size of this cloud is characterized by a ``cloud parameter'' $b$
that is equal to the sum of the separations $|{\bf x}_a-{\bf
x}_{a+1}|$ between the monopoles.  Any value $b \ge r \equiv |{\bf
x}_1-{\bf x}_{N+1}|$, the separation of the two massive monopoles, is
possible, and the static energy is independent of $b$.  All other
trace of the massless monopole positions appears to be lost in this
limit.  In particular, the fields show no special behavior at the
${\bf x}_a$ corresponding to the massless monopoles.  In fact, any
change in these ``positions'' that leaves $b$ invariant can be
compensated by a gauge transformation of the solution.

Our goal in this paper is to gain further insight into the nature of
these massless monopoles by examining the role that the cloud plays in
monopole scattering.  To do this we use the moduli space
approximation \cite{msa}, which reduces the dynamics of the infinite
number of
field degrees of freedom to that of a finite number of collective
coordinates $q_i$.  The Lagrangian for this reduced set of variables 
can be written as 
\begin{equation}
    L = {1\over 2} g_{ij}(q) \dot q^i \dot q^j
\end{equation}
where 
\begin{equation}
   {\cal G} = g_{ij}(q) dq^i dq^j
\end{equation}
is a naturally defined metric on the moduli space of solutions.  The
evolution of the collective coordinates is simply geodesic motion in
this metric.  When the monopoles are all massive, this reduction to a
finite number of degrees of freedom fits quite well with the point of
view that particles arising from classical soliton solutions should be
seen as having a similar status to those arising as quanta of the
elementary fields, and that the solutions with higher charge should be
interpreted as multiparticle states.  When some of the monopoles
become massless, the classical solutions lose their manifest
multiparticle nature.  Nevertheless, the massless monopole collective
coordinates survive (although with their physical interpretations
modified) and the moduli space Lagrangian has a smooth
limit~\cite{kwynonabelian}, provided that the total magnetic charge is
Abelian.

The moduli space approximation is usually expected to be valid in the
limit of small monopole velocities.  This ensures that the monopole
kinetic energy is small enough that the excitation of the
nonzero-frequency modes of the massive fields is energetically
suppressed.  Although simple energetic arguments are not sufficient to
rule out 
excitation of any massless fields that might be present, more
detailed analysis \cite{validity} shows that the approximation holds
if the long-range fields are all Abelian.  As we will see, the
situation with massless non-Abelian fields is more complex.

To use the moduli space approximation, we need the metric on the space
of multimonopole solutions.  The exact form of this metric is 
known for only a relatively few cases.  However, these include
\cite{kwy,proofs} the case where there is at most one of each species
of fundamental monopole, which is precisely what we need.
Unfortunately, the form of this metric that was obtained in
Ref.~\cite{kwy} was expressed in terms of the positions and U(1)
phases of the $(N+1)$ individual monopoles.  Although these are the
natural coordinates to use when the monopoles are all massive, they
are much less useful in the limit we are interested in, where the
massless monopole positions have little direct physical meaning.  A
more natural set of coordinates would include the positions and U(1)
phases of the two massive monopoles, the cloud parameter $b$, and the
parameters needed to specify the global SU($N$) orientation of the
cloud.

Thus, our first task is to change to a new set of variables.  We do
this in two steps.  In Sec.~II we transform to an intermediate set of
coordinates that allows us to separate the metric into two parts, one
that gives the terms in the Lagrangian containing $\dot r$ and $\dot
b$, and an ``angular'' part that contains the entire dependence on the
spatial and gauge orientation angles and phases.  The natural next
step would be to relate these intermediate coordinates to the spatial
Euler angles and their gauge counterparts.  This turns out to be
somewhat nontrivial.  However, we can bypass this step by rephrasing
the problem in terms of the corresponding angular momenta and charges.
We do this in Sec.~III, leaving the discussion of
the Euler angles to the Appendix.  With these results in hand, we are
then ready to determine the geodesics of the metric and the behavior
of the monopoles in scattering processes.  This is described in
Sec.~IV.  We summarize our results and make some concluding remarks in
Sec.~V.

\section{The moduli space metric}

The moduli space of solutions for a system comprising two
distinct fundamental monopoles was obtained in Ref.~\cite{twomono}; the
extension to any number of distinct monopoles was conjectured in
Ref.~\cite{kwy} and proven in Ref.~\cite{proofs}.  Locally,
this space can be written as a product of a flat 4-dimensional space,
corresponding to the center-of-mass position and an overall U(1)
variable, and a relative moduli space.   For the case of interest to
us, an SU($N+2$) theory with one monopole of each species, the moduli
space is $4(N+1)$-dimensional, and the $4N$ coordinates for the relative
moduli space can be chosen to be the relative positions ${\bf r}_A =
{\bf x}_A - {\bf x}_{A+1}$ ($A=1,2, \dots, N$) and the corresponding
relative U(1) phases $\psi_A$.  

In Ref.~\cite{kwynonabelian} it was shown that when the $SU(N+2)$
symmetry is broken to U(1)$\times$SU($N$)$\times$U(1) this metric can
be written as
\begin{equation}
     {\cal G}_{\rm rel}  = \mu \left[\sum_A  d{\bf r}_A \right]^2
     +\kappa \sum_A  \left[\frac{1}{r_A}
      d{\bf r}_A^2 + r_A (d\psi_A+\cos\theta_Ad\phi_A)^2\right]
    - \lambda
       \left[\sum_A r_A\,(d\psi_A+\cos\theta_Ad\phi_A)\right]^2
\label{rAmetric}
\end{equation}
where $r_A$, $\theta_A$, and $\phi_A$ are the spherical coordinates
of the vector ${\bf r}_A$, $\mu$ is the reduced mass of the two
massive monopoles, and
\begin{eqnarray}
     \kappa &=& {g^2 \over 8\pi}   \cr
     \lambda &=& {g^2 \mu \over g^2 + 8 \pi \mu b} 
       = \mu \left[{r_c \over 2b + r_c} \right]  \, \,.
\end{eqnarray}
Here $g=4\pi/e$ (with $e$ the gauge coupling of the
theory) is the magnitude of the fundamental
monopole magnetic charge, while $r_c \equiv 2 \kappa/\mu$ is approximately
equal to the sum of the core radii of the two massive monopoles.

The difficulty with using this expression for our purposes is
that, as we noted in Sec.~I, the positions of the massless monopoles
do not have any direct physical meaning.  We would therefore like to
transform to a more physically meaningful set of variables.  This set
should include the cloud parameter $b$ and the spherical coordinates
$r$, $\theta$, and $\phi$ of the vector $\bf r$ that gives the
relative separation of the two massive monopoles, but must also
include the relative U(1) phase and a number of SU($N$) orientation
angles.

One might expect that $N^2-1 = {\rm dim}\,[{\rm SU}(N)]$ such angles
would be needed to specify the gauge orientation of the solution.  For
$N>2$ this would require more collective coordinates than are
available.  However, examination of the explicit solutions shows that
there is always a U($N-2$) subgroup of the unbroken SU($N$) that leaves
the fields invariant [i.e., any of these SU($N+2$) solutions is
essentially an embedding of an SU(4) solution].  Hence, there are only 
$4N-5 = {\rm dim}\,[{\rm SU}(N)/{\rm U}(N-2)]$ global SU($N$)
parameters.  Together with the U(1) phases, the cloud size $b$, and the
positions of the two massive monopoles, this gives a total of $4(N+1)$
collective coordinates, as required.

We begin by defining complex
coordinates\footnote{In the notation of Ref.~\cite{kwynonabelian}, $z_A^1
=
\xi_A/2$ and $z_A^2 = \zeta_A^*/2$.}
\begin{eqnarray}
   z_A^1 &=& \sqrt{r_A}  \cos(\theta_A/2)
       e^{-i\,(\phi_A+\psi_A)/2} 
       \nonumber\\
   z_A^2 &=& \sqrt{r_A} \sin(\theta_A/2)
       e^{i\,(\phi_A-\psi_A)/2}   \, \,.
\end{eqnarray}
These satisfy
\begin{equation}
   {\bf r}_A = \bar z_A^a (\bsigma)_{ab} z^b_A
\end{equation}
where $\bar z_A^a =(z_A^a)^*$ and the $\sigma_k$ are the Pauli
matrices.  Hence, the relative position of the
two massive monopoles is\footnote{Omitted indices on the $z_A^a$ and
related quantities should be understood as being summed over.}
\begin{equation}
    {\bf r}= \bar z  \bsigma z 
\label{rFROMz}
\end{equation}
while
\begin{equation}
    b = \bar z z \, \,.
\label{bFROMz}
\end{equation}
When rewritten in terms of these coordinates, the metric takes the
form 
\begin{eqnarray} 
       {\cal G}_{\rm rel} &=& 
           \mu \,[\bar z\, \bsigma\, dz + d\bar z \,\bsigma\, z]^2
      + 4 \kappa \, d\bar z\, dz
      + \lambda\, [\bar z\, dz -d\bar z\, z]^2 \cr
     &=& \mu \,d{\bf r}^2  + 4 \kappa\, d\bar z\, dz + 
        \lambda\, [\bar z\, dz -d\bar z\, z]^2  \, \,.
\label{zmetric}
\end{eqnarray}

Next, we find an SU(2) matrix $U$ satisfying 
\begin{equation}
       U \hat {\bf r}\cdot \bsigma U^{-1} = \sigma_3
\label{Uequation}
\end{equation}
where $\hat{\bf r} = {\bf r}/r$. (Thus, $U$ corresponds to a
rigid rotation of the entire assembly of monopoles that puts the two
massive monopoles on the $z$-axis.)  This allows us to define a new
set of complex variables 
\begin{equation}
     w_A^a = U_{ab} z_A^b
\label{ZtoW}
\end{equation}
in terms of which $r$ and $b$ are given by the simple expressions
\begin{equation}
      \bar w \sigma_j w = r \delta_{j3}  \, , \qquad \bar w w =b 
 \, \,.
\label{wIdentities}
\end{equation}

These identities imply that the $w^a_A$ must be of the form
\begin{eqnarray}
     w_A^1 &=& \sqrt{b+r \over 2} p_A^{(1)} \cr
     w_A^2 &=& \sqrt{b-r \over 2} p_A^{(2)}
\label{wform}
\end{eqnarray}
where the $p_A^{(a)}$ are a pair of complex $N$-component vectors
obeying 
\begin{equation}
       \bar p_A^{(a)} p_A^{(b)} = \delta_{ab} \, \,.
\end{equation}
For later use, we also introduce $N-2$ orthonormal vectors
$e^{(r)}_A$ ($r = 3, 4, \dots, N$) orthogonal to the $p_A^{(a)}$.

We must now rewrite the metric in terms of these new variables.  It
will be helpful to introduce some new notation.  First, we define two
unit vectors $\hat {\bf n}_1$ and $\hat {\bf n}_2$ orthogonal to $\hat
{\bf r}$ by requiring that 
\begin{equation}
       U \hat {\bf n}_a\cdot \bsigma U^{-1} = \sigma_a \, \,.
\label{nhatdef}
\end{equation}
Next, we define $A_j$ and $V_j$ ($j = 1, 2, 3$) by
\begin{equation}
     dU\, U^{-1} =   -{1\over 2i} A_j \sigma_j
\end{equation}
and 
\begin{equation}
   V_j = i [ \bar w \,\sigma_j \,dw - d\bar w \,\sigma_j\, w] \, \,.
\label{Vjdef}
\end{equation}
Finally, we define
\begin{equation}
   V_0 = i [ \bar w \,dw - d\bar w \, w] \, \,.
\label{V0def}
\end{equation}

We also need two identities.  First, by differentiating both sides of
Eq.~(\ref{Uequation}), one is led to
\begin{equation}
   d{\bf r} = \hat {\bf r}\, dr + r ( \hat {\bf n}_1 A_2 
    - \hat {\bf n}_2 A_1) \, \,.
\end{equation}
Next, viewing the $w^a$ as vectors in an $N$-dimensional complex
vector space, we write
\begin{equation}
    d\bar w \,dw = d\bar w^a_A \,[ \Pi^L_{AB} + \Pi^T_{AB}]
             \, dw_B^a
\label{dwdw}
\end{equation}
where 
\begin{equation}
     \Pi^L_{AB} = \sum_{a=1}^2 p_B^{(a)} \, \bar p_A^{(a)}
\end{equation}
is the projection operator onto the two-dimensional
subspace spanned by $w^1$ and $w^2$ (or by $z^1$ and $z^2$) and
\begin{equation}
     \Pi^T_{AB} = \sum_{r=3}^N  e^{(r)}_A \, \bar e^{(r)}_B   \, \,.
\end{equation}
is the projection operator onto the orthogonal
$(N-2)$-dimensional subspace.
Using Eqs.~(\ref{wIdentities}), (\ref{wform}),
(\ref{Vjdef}), and (\ref{V0def}), we obtain the second identity,
\begin{equation}
   d\bar w\, dw = { [d(b+r)]^2 \over 8 (b+r) }
               + { [d(b-r)]^2 \over 8 (b-r) }
          + {b \over 4 (b^2-r^2)} \sum_{\nu=0}^3 V_\nu^2 
           - {r \over 2(b^2-r^2) }(V_0 V_3)
          + d\bar w \,\Pi^T \,dw   \,\, .
\end{equation}

We have not yet obtained explicit expressions for $\theta$,  
$\phi$, or the gauge orientation phases in terms of the
$w_A^a$  or $U$.  However, it is easy to see that neither the $A_i$ and
$V_\nu$ nor the combination $d\bar w \,\Pi^T \,dw$ can contain any
factors of $dr$ or $db$.   With this in mind, and using our two
identities and the above definitions, we substitute
Eq.~(\ref{ZtoW}) into Eq.~(\ref{zmetric}) and obtain
\begin{equation}
    {\cal G}_{\rm rel}
        = \mu \, dr^2 + \kappa \, \left\{ {[d(b+r)]^2 \over 2 (b+r) }
               + { [d(b-r)]^2 \over 2 (b-r) } \right\}
   + {\cal G}_{\rm ang}
\label{metricdecomp}
\end{equation}
where 
\begin{eqnarray}
    {\cal G}_{\rm ang} 
        &=& \sum_{i=1}^2 \left[ \mu \, r^2 A_i^2 
          + {\kappa b \over b^2-r^2}\, V_i^2 
          + \kappa \,(2 A_i V_i + b A_i^2) \right] 
       + \left({\kappa b \over b^2-r^2} - \lambda \right)\,
            (V_0 + r A_3)^2  
     \cr  &&\qquad
       + {\kappa b \over b^2-r^2}\, (V_3 + bA_3)^2
        - {2\kappa r \over b^2-r^2}\, (V_0 + r A_3)(V_3 + bA_3)
          + 4  \kappa \, d\bar w \,\Pi^T \,dw  
\label{GangAV}
\end{eqnarray}
contains the entire dependence of the metric on the spatial angles
and the various gauge orientation phases.   
In the next section we will rewrite ${\cal G}_{\rm ang}$ in a
form that will allow us to analyze the evolution of these angle and gauge
variables.  

Before doing so, we comment on a point that we have skipped over.
Equation~(\ref{Uequation}) does not completely determine the matrix
$U$, but instead leaves the freedom to make a further transformation
of the form
\begin{equation}
        U \longrightarrow  e^{i\chi \sigma_3/2}\,U \, \,.
\end{equation}
This implies a corresponding transformation 
\begin{equation}
      w \longrightarrow  e^{i\chi \sigma_3/2} \, w
\end{equation}
of the $w_A^a$, and rotates the unit vectors $\hat{\bf n}_1$ and 
$\hat{\bf n}_2$ so that 
\begin{equation}
    \hat{\bf n}_1  + i \hat{\bf n}_2 \longrightarrow
      e^{-i \chi} (\hat{\bf n}_1  + i \hat{\bf n}_2) \, \,.
\end{equation}

This ambiguity in the definition of $U$ is not simply due to our
failure to impose a sufficient number of conditions.  Instead, it 
reflects the fact that these solutions possess an axial symmetry.
They are invariant under rotations about the axis joining the two
massive monopoles, provided that these are combined with an appropriate
global gauge transformation.
Because of
this symmetry, there is no natural way to pick out any particular pair
of axes in the plane perpendicular to $\hat{\bf r}$.  As a result, the
form of the metric should be invariant under such redefinitions of
$U$.   To verify that it is, we note that the auxiliary quantities $A_i$
and $V_\nu$ transform as  
\begin{eqnarray}
      A_1 + i A_2 &\longrightarrow&  e^{-i\chi} ( A_1 + i A_2 )   \cr
      A_3 &\longrightarrow&  A_3 + d\chi   \cr
      V_1 + i V_2 &\longrightarrow& e^{-i\chi} ( V_1 + i V_2 )   \cr
      V_3 &\longrightarrow&  V_3 - b\, d\chi \cr
      V_0 &\longrightarrow&  V_0 - r\, d\chi \, \,.
\end{eqnarray}
Substituting these results into Eq.~(\ref{GangAV}), we see that the 
metric is indeed unchanged, thus providing a useful consistency check on
our calculations.

\section{Angles, phases, and conserved quantities}

In this section we will bring Eq.~(\ref{GangAV}) for ${\cal G}_{\rm
ang}$ into a form more suitable for analyzing the dynamics of the
angle and phase variables.  The most straightforward method for doing
this would begin by defining generalized Euler angles to describe the
spatial and the U($N$) phase orientations.  Given these, one can
construct a set of 1-forms $\omega^j$ invariant under both the spatial
SO(3) and the internal SU($N$) symmetries of the theory, and write
\begin{equation} 
     {\cal G}_{\rm ang} = I_{ij}(r,b) \omega^i \omega^j \, \,.
\label{Gomega}
\end{equation}
The symmetries of the system imply the existence of a conserved angular
momentum and conserved U($N$) charges.  The ``body-frame'' components of
these quantities, defined by
\begin{equation}
       X_i\, dt = I_{ij}(r,b) \omega^j
\end{equation}
are the variables conjugate to the invariant one-forms.  Writing 
\begin{equation}
    {\cal G}_{\rm ang} = (I^{-1})^{ij}(r,b) X_i X_j dt^2 
\label{GwithIinv}
\end{equation}
and then substituting this into Eq.~(\ref{metricdecomp}) gives a
convenient starting point for studying the dynamics of the system.

A complication in our case is that, because of the extra symmetries of
these multimonopole solutions, the number of angular variables is less
than might be expected.  For most choices of the magnetic charges, an
arbitrary configuration would require 3 spatial Euler angles and $N^2$
phases specifying the U($N$) orientation.  By contrast, the relative
moduli space in our case is $4N$-dimensional so that, after extracting
$r$ and $b$, we are left with only $4N -2$ angle and phase variables.
The ``missing'' $(N-2)^2 + 1$ variables are explained by extra
invariances of the solutions we are considering.  First, we noted
previously, these SU($N+2$) solutions are essentially equivalent to
embeddings of SU(4) solutions, so that there is always a U($N-2$)
subgroup that leaves the configuration invariant.\footnote{Note,
however, that the particular U($N-2$) subgroup depends on the SU($N$)
orientation of the configuration, and thus changes as the various
phases vary with time.} The $(N-2)^2$ ``Euler angles'' corresponding
to this subgroup play no role in the dynamics, and hence do not enter
the Lagrangian. Second, the axial symmetry discussed at the end of the
previous section implies that some linear combination of a spatial
Euler angle and a U($N$) phase must also be absent from the
metric.\footnote{The apparent analogy between a rigid body symmetric
top and our axially symmetric field configurations is a bit
misleading.  In the former case, time-dependent rotations about the
symmetry axis are physically meaningful and have kinetic energy
associated with them.  Although the rotation angle about this axis does
not appear in the Lagrangian, its time derivative does.  In our
case, such rotations, whether time-dependent or not, are undetectable
and have no associated kinetic energy.  Thus, the best analogue is not
an ordinary symmetric top with principle moments of inertia $I_1 =I_2
\ne I_3 \ne 0$, but rather an infinitely thin top with $I_1 =I_2$ and
$I_3=0$.}

The fact that these extra symmetries involve mixtures of spatial
rotations and gauge transformations complicates the explicit
construction of the $\omega^i$ appearing in Eq.~(\ref{Gomega}); they
are not simply subsets of the standard invariant one-forms for SO(3)
and U($N$).  Fortunately, it turns out that we can bypass the
construction of the $\omega^i$ and go directly from Eq.~(\ref{GangAV})
to an expression of the form given in
Eq.~(\ref{GwithIinv}).  We will follow this procedure in this section,
leaving the discussion of the generalized Euler angles and the
$\omega^i$ to the Appendix.

We begin by identifying the conserved quantities of the
system.  Recall that if a metric has a Killing vector of the form
\begin{equation}
   K_{(i)} = K_{(i)}^a {\partial \over \partial q^a} 
\label{Killingdef}
\end{equation}
then the charge
\begin{equation}
   Q_{(i)}= K_{(i)}^a \dot q_a = g_{ab} K_{(i)}^a \dot q^b
\label{chargedef}
\end{equation}
is a constant along any geodesic.  Since the classical solutions of
the moduli space Lagrangian are all geodesics, these $Q_{(i)}$ are the
desired conserved charges.

The Killing vectors of the metric of Eq.~(\ref{rAmetric}) were
determined in Ref.~\cite{kwynonabelian}. Written in terms of our
complex variables $z_A^a$, the rotational Killing vectors take the
form
\begin{equation}
    L_k = {i\over 2} 
      \left[ z_A^b \, (\sigma_k)_{ab}\, {\partial\over \partial z_A^a}
   - \bar z_A^b\, (\sigma_k)_{ba} \,{\partial\over \partial \bar z_A^a}
          \right]
\label{rotKilling}
\end{equation}
while the $N^2$ Killing vectors
\begin{equation}
    E_{AB} = {i \over \sqrt{2}}
     \left[ z_A^a \,{\partial\over \partial z_B^a}
     -  \bar z_B^a\, {\partial\over \partial \bar z_A^a}\right]
\label{UNKilling}
\end{equation}
satisfy the algebra of U($N$).  

Using Eqs.(\ref{Killingdef}) and (\ref{chargedef}), we construct from
these the angular momentum
\begin{eqnarray}
     {\bf J}\,dt  &=& \mu \, (\bar z  \,\bsigma  \,z) {\bf \times}
        [ \bar z \, \bsigma\, dz + d \bar z \, \bsigma \, z]
           + i \kappa \, [ \bar z \, \bsigma\, dz  
           - d \bar z \, \bsigma \, z]
         - i \lambda  \,[\bar z \, dz - d\bar z \, z]
             (\bar z  \,\bsigma  \,z)  \cr
       &=& \mu \, {\bf r} {\bf \times} d{\bf r}
         + \kappa \, \left[ i(\bar w \, U \bsigma  U^{-1} \, dw
           - d\bar w \, U  \bsigma  U^{-1} \, w )
           + i \bar w ({\cal A}  \, U \, \bsigma \, U^{-1}
               +  U \bsigma  U^{-1} \,{\cal A})w \right]  
        \cr &&\qquad
        - \lambda \, {\bf r}\, (V_0 + rA_3)
\label{Jequation}
\end{eqnarray} 
(where ${\cal A}= A_j\sigma_j$) and the U($N$) charges 
\begin{eqnarray}
      {\cal E}_{AB}\,dt &=&  i \sqrt{2}\kappa \left[ d\bar z_B \, z_A
         -  \bar z_B \, dz_A \right]
        +   i \lambda\, \bar z_B \, z_A  \,[\bar z \, dz - d\bar z \, z]
      \cr
      &=& \sqrt{2}\kappa \left[ i(d\bar w_B \, w_A
         -  \bar w_B \, dw_A - \bar w_B \,\sigma_j \, w_A \, A_j
                  \right] 
            +\sqrt{2}  \lambda \, \bar w_B \, w_A (V_0 + rA_3)  \, \,.
         \cr  &&
\label{Eabequation}
\end{eqnarray}

The components implied by the bold-face notation in
Eq.~(\ref{Jequation}) and by the subscripts in Eq.~(\ref{Eabequation})
are those corresponding to a fixed ``space-frame'' and are conserved.
What we actually need are the components corresponding to a moving
``body-frame''.  For the angular momentum, this requires identifying
three axes that rotate with the monopole system; a suitable choice
is $\hat {\bf n}_1$, $\hat {\bf n}_2$, and $\hat {\bf n}_3
\equiv \hat{\bf r}$.  The body frame components (which are themselves
rotational scalars) are then
\begin{equation}
    J_i\, dt \equiv \hat {\bf n}_i \cdot {\bf J} \,dt = \cases{ (\mu r^2 +
           \kappa b) A_i + \kappa V_i \, & $i = 1, 2$ \cr \cr
    - r  \lambda (V_0 + rA_3) + \kappa (V_3 + b A_3) \, & $i = 3 $
       \, \,.}          
\label{bodyJ}
\end{equation}
In obtaining this, we have used Eqs.~(\ref{nhatdef}) and (\ref{Vjdef}).

To construct a ``body-frame'' for the U($N$) charges, we need a set of
$N$ complex basis vectors in the internal space that transform under
the fundamental representation under the action of the ${\cal
E}_{AB}$.  We can take these to be the $p_A^{(a)}$ and the $e^{(r)}_A$
that were introduced in Sec.~2.
Using the $p_A^{(a)}$ and the Pauli matrices, we can construct three
U($N$)-invariant charges
\begin{equation}
    T_k = {1 \over \sqrt{2}} \bar p_A^{(a)}\, [\tau_k]_{ab} \,p_B^{(b)} 
                 \,{\cal E}_{AB}  \, \,.
\end{equation}
(The definition of these has been chosen so that they correspond to
generators of an SU(2) subgroup with the standard normalization.)
Straightforward calculations give
\begin{equation}
    T_i\, dt  = \cases{ \kappa
         \left[ {b \over \sqrt{b^2 - r^2}} V_i 
              + \sqrt{b^2 - r^2} A_i \right] \,
         & $i = 1, 2$  \cr  \cr
          r \lambda (V_0 +  rA_3) - \kappa (V_3 + b A_3) \,
            & $i = 3 $  \, \,.}
\label{bodyT}
\end{equation}
Comparing Eqs.~(\ref{bodyJ}) and (\ref{bodyT}), we see that 
\begin{equation}
      J_3 + T_3 =0
\end{equation}
which is a reflection of the axial symmetry of the solutions. 

Additional ``body-frame'' components can be constructed by utilizing
contractions with the $e^{(r)}_A$.  Components involving two such
contractions vanish; i.e.,\footnote{This is just a restatement of the
fact that $(N-2)^2$ eigenvalues of the moment of inertia tensor vanish
because of invariance under a U($N-2$) subgroup.}
\begin{equation}
       \bar e^{(r)}_A \,  e^{(s)}_B \,  {\cal E}_{AB} =0 \, \,.
\label{eEe}
\end{equation}
On the other hand, the charges 
\begin{equation}
    t_{ar} =   \bar p^{(a)}_A \, e^{(r)}_B \,
       {\cal E}_{AB}
\end{equation}
and their complex conjugates $\bar t_{ar}$ are in general nonzero;
they are related to the last term in Eq.~(\ref{GangAV}) by 
\begin{equation}
       {d\bar w \over dt}\,\Pi^T \,{dw  \over dt} = {1 \over \kappa^2} 
      \sum_{r=3}^N \left[ {t_{1r} \bar t_{1r} \over b+r }
         + {t_{2r} \bar t_{2r} \over b-r }\right] \, \,.
\label{dbarwPIdw}
\end{equation}

Finally, we define a conserved relative U(1) charge
\begin{eqnarray}
       Q\,dt &=& {1 \over \sqrt{2}} \,{\cal E}_{AA} \,dt 
        = {1 \over \sqrt{2}} \left[ \Pi_{AB}^T
                 + \Pi_{AB}^L \right]  \,{\cal E}_{AB} \,dt
        = {1 \over \sqrt{2}}  \sum_{a=1}^2 p_B^{(a)} \, \bar p_A^{(a)} 
                \, {\cal E}_{AB} \,dt \cr
        &=& (b \lambda - \kappa) (V_0 + r A_3) \, \,.
\label{Qdef}
\end{eqnarray}

Using Eqs.~(\ref{GangAV}), (\ref{bodyJ}), (\ref{bodyT}),
(\ref{dbarwPIdw}), and (\ref{Qdef}), we can now rewrite ${\cal G}_{\rm
ang}$ in terms of the body-frame components, obtaining
\begin{eqnarray}
    {\cal G}_{\rm ang}  &=& \left\{ {1 \over r^2(\kappa + \mu
b)}\sum_{i=1}^2
    \left[ b J_i^2 + \left(b +{\mu r^2 \over \kappa} \right) T_i^2
     -2\sqrt{b^2 -r^2} J_i T_i \right]  \right.
     \cr &+&    \left.
       {1 \over \kappa(b^2 -r^2)} \left[ b (J_3^2 +Q^2)
       + 2 r J_3 Q \right] 
      + {\mu \over \kappa^2}Q^2
     + {4 \over \kappa} \sum_{r=3}^N \left[ {t_{1r} \bar t_{1r} \over b+r
}
         + {t_{2r} \bar t_{2r} \over b-r }\right]  \right\} dt^2 \, \,.
\label{Gangfinal}
\end{eqnarray}

It must be remembered that the ``body frame'' components that appear
in this expression are not in general conserved.  While the square of
the angular momentum,
\begin{equation}
      {\bf J}^2 = J_1^2 + J_2^2 + J_3^2
\label{Jsquare}
\end{equation}
the SU($N$) Casimir, 
\begin{equation}
      T^2 = T_1^2 + T_2^2 + T_3^2 
   + \sum_{r=3}^N  \left(t_{1r} \bar t_{1r} + t_{2r} \bar t_{2r}\right)
\label{Tsquare}
\end{equation}
and the U(1) charge $Q$ are constant, determining the time-dependence
of the individual components of $\bf J$ and $T$ is in general rather
complex.  There is considerable simplification, however, if the
SU($N$) charges all vanish.  The only nonzero charges remaining are
then $J_1$, $J_2$, and $Q$, with the former two entering ${\cal
G}_{\rm ang}$ only through ${\bf J}^2 = J_1^2 +J_2^2$; using the
relation between $V_i$ and $A_i$ that follow from the vanishing of the
$T_i$, one finds that this is given by
\begin{equation}
     {\bf J}^2 = \left(\mu + {\kappa \over b}\right)^2 r^4
       (\dot\theta^2 + \sin^2 \theta\, \dot\phi^2) \, \,.
\end{equation}
Equation~(\ref{Gangfinal}) then reduces to 
\begin{equation}
    {\cal G}_{\rm ang}(T=0) = \left\{
      { b {\bf J^2}\over  r^2(\kappa + \mu b)} 
       + \left[{\mu \over \kappa^2} + {b \over \kappa(b^2 -r^2)}
       \right]  Q^2   \right\} dt^2 \, \,.
\label{GangTzero}
\end{equation}

\section{Monopole trajectories}

We are now ready to obtain the equations of motion of our system and
study the behavior of the cloud and the massive monopoles in
scattering processes.  We restrict ourselves here to the case of
vanishing SU($N$) charges, $T=0$, leaving a discussion of the more
general case to the Appendix.  

Because the massive monopoles have well-defined positions while the
massless ones do not, there may be a tendency to view the various
scattering processes as resulting from interactions between the
massive monopoles, with the cloud dynamics being in the background.
This would be incorrect.  The massive monopoles correspond to two
mutually commuting SU(2) subgroups of SU($N+2$).  Were it not for the
presence of the cloud, they would not interact at all and would move
on straight lines with constant velocity.  Hence, when following the
motion of the massive monopoles it is important to also keep track of
the size and motion of the cloud at the same time.

Equations~(\ref{metricdecomp}) and (\ref{GangTzero}) provide a
suitable starting point for obtaining the equations of motion of our
system.  Because the angle and phase variables enter the latter only
through conserved quantities, we can view ${\cal G}_{\rm ang}$ as
defining an effective potential for $r$ and $b$ that can be combined
with the first two terms in Eq.~(\ref{metricdecomp}) to yield an
effective Lagrangian\footnote{We omit the term $(\mu/\kappa^2)
Q^2$, which gives a $Q$-dependent increase in energy but has no effect
on the evolution of $b$ or $r$.}
\begin{equation}
    L_{\rm eff} = {\mu \over 2}\dot r^2  
       + {\kappa\over 4} \left[{(\dot b+ \dot r)^2 \over b+r}
             + {(\dot b - \dot r)^2 \over b-r} \right]  
     - { b {\bf J}^2\over 2 r^2(\kappa + \mu b)}
       - {b Q^2\over 2\kappa(b^2 -r^2)} \, \,.
\label{Leffective}
\end{equation}

We begin our analysis of the solutions of this Lagrangian by
examining the asymptotic solutions at large $r$ and $b$; because the
terms containing $\bf J$ and $Q$ both fall rapidly in this regime, it 
is sufficient to do this asymptotic analysis with ${\bf J}=Q=0$.   We
find that the cloud and the massive monopoles decouple from each
other at large times.  After
studying the properties of these asymptotic solutions, we then turn to
the behavior at finite time, where the actual interactions take place,
beginning with ${\bf J}=Q=0$.  Finally, we consider the effects of
nonzero $\bf J$ and $Q$.

Before beginning this analysis, we must address the apparent
singularity in the kinetic energy at $r=b$.  This singularity is not
physical, and can be eliminated by defining
\begin{equation}
   \matrix{ x = \sqrt{\kappa} \left[ \sqrt{b+r} + \sqrt{b-r}\right]
\cr\cr
    y = \sqrt{\kappa} \left[ \sqrt{b+r} - \sqrt{b-r}\right] }
                \, \qquad\qquad  0 \le y \le x \, \,.
\label{firstoctant}
\end{equation}
In terms of these coordinates, the Lagrangian takes the nonsingular
form
\begin{equation}
    L  = {1 \over 2} (\dot x^2 + \dot y^2) 
      +  {1 \over 2a^2}  (x \dot y + y \dot x)^2
      - {a^2 (x^2 +y^2)  \over 2 x^2y^2 (a^2 + x^2 +y^2)}{\bf J}^2
      - {2 (x^2 +y^2) \over (x^2-y^2)^2 } Q^2
\label{xyLag}
\end{equation}
where 
\begin{equation}
      a = {2 \kappa \over \sqrt{\mu}} \, \,.
\label{adef}
\end{equation}
As we have indicated in Eq.~(\ref{firstoctant}), the entire physical
range $0 \le r \le b < \infty$ is mapped onto the octant $0 \le y \le
x$ of the $x$-$y$ plane.  Nevertheless, it will become clear that if
either $\bf J$ or $Q$ vanish some trajectories can cross the boundaries of
this octant.  These can be understood as follows.  The $x$-axis
corresponds to $r=0$.  A trajectory crossing this line corresponds to a
motion in which the two massive monopoles approach head-on, meet (at
$y=0$), and then pass through one another. To accommodate this, we
extend the definition of $x$ and $y$ into the next octant by
\begin{equation}
   \matrix{ x = \sqrt{\kappa} \left[ \sqrt{b-r} + \sqrt{b+r}\right]
    \cr\cr 
    y = \sqrt{\kappa} \left[ \sqrt{b-r} - \sqrt{b+r}\right] }
     \, \qquad\qquad  0 \le -y \le x \, \,.
\end{equation}
Similarly, a trajectory crossing the line $x=y$ corresponds to a motion in
which the cloud shrinks to its minimum size, $b=r$, and then begins to
grow again.  To describe this, we define 
\begin{equation}
    \matrix{x=  \sqrt{\kappa} \left[ \sqrt{b+r} - \sqrt{b-r}\right]
    \cr\cr 
    y =  \sqrt{\kappa} \left[ \sqrt{b+r} + \sqrt{b-r}\right] }
           \, \qquad\qquad  0 \le x \le y \, \,.
\end{equation}
Extending these definitions in a similar fashion to the other octants,
we obtain an eight-fold mapping of the physical $r$-$b$ space onto the 
$x$-$y$ plane.  Throughout the plane we have the relations
\begin{eqnarray}
    r &=&  {1 \over 2 \kappa} \,|xy|   \cr &&\cr
    b &=&  {1 \over 4 \kappa} \, (x^2 + y^2)
\end{eqnarray}
so that the $x$- and $y$-axes correspond to $r=0$ while the lines
$y=\pm x$ correspond to minimum cloud size, $b=r$.

\subsection{Asymptotic solutions for ${\bf J}=Q=0$}

If ${\bf J}=Q=0$ the Lagrangian of Eq.~(\ref{xyLag}) reduces to 
\begin{equation}
    L = {1 \over 2} (\dot x^2 + \dot y^2) 
      +  {1 \over 2a^2}  (x \dot y + y \dot x)^2 \, \,.
\label{rbLag}
\end{equation}
The resulting Euler-Lagrange equations are
\begin{eqnarray}
    0 &=&  (a^2  +y^2) \ddot x + xy \ddot y+ 2y \dot x \dot y    \cr
    0 &=&  (a^2 + x^2) \ddot y + xy \ddot x+ 2x \dot x \dot y  \, \,.
\label{ELeqs}
\end{eqnarray}
Multiplying the first of these by $x$ and the second by
$y$, and then subtracting, we find that $ x \ddot x = y \ddot y$.
Using this to rewrite Eq.~(\ref{ELeqs}), we obtain 
\begin{eqnarray}
    0 &=&  (a^2 + x^2 +y^2) \ddot x + 2y \dot x \dot y    \cr
    0 &=&  (a^2 + x^2 +y^2) \ddot y + 2x \dot x \dot y  \, \,.
\label{newELeqs}
\end{eqnarray}

There are two constants of the motion.  The time-independence of the
Lagrangian implies the conservation of the energy (which is actually
equal to the Lagrangian)
\begin{equation}
     E = {1 \over 2} (\dot x^2 + \dot y^2) 
      +  {1 \over 2a^2} (x \dot y + y \dot x)^2 \, \,.
\label{xyenergy}
\end{equation}
In addition, multiplication of the first of Eq.~(\ref{newELeqs})
by $\dot y$ and the second by $\dot x$ shows that 
\begin{equation}
     B = {1 \over  a^2}\dot x \dot y (a^2 + x^2 +y^2)
\label{Bdef}
\end{equation}
is also conserved.  (We have no deep understanding as to why $B$ is
constant; we have not found any generalization of $B$ that is 
conserved when either $\bf J$ or $Q$ is nonzero.)  

If $B \ne 0$, neither $\dot x$ nor $\dot y$ can vanish, implying that
both $x$ and $y$ vary monotonically; if $B=0$ but $E \ne 0$, then one of
$\dot x$ or $\dot y$ vanishes for all $t$, while the other never
vanishes.  Hence, there are no closed orbits in the $x$-$y$ plane. It is
then straightforward to show that $x^2 + y^2$ must tend to infinity as
$t \rightarrow \pm\infty$. 
There are two possibilities for the large time behavior.   The
first is that $dy/dx$ tends to a nonzero finite constant.  
Examination of Eq.~(\ref{newELeqs}) shows that this is possible only
if $dy/dx \rightarrow \pm 1$, which can only happen if
$E=\pm B$.  This gives solutions where $x^2 = y^2$ varies linearly with
time, corresponding to a solution with a minimal ($b=r$) cloud
surrounding two massive monopoles with constant relative velocity.  

The other possibility is that $|dy/dx|$ tends to either 0 or $\infty$;
because of the eight-fold mapping of $b$ and $r$ onto the $x$-$y$
plane, the two are are equivalent.  We will analyze in detail the case
where $dy/dx \rightarrow 0$, so that asymptotically $|x| \gg |y|$.  By
integrating the dominant terms in the second of Eq.~(\ref{newELeqs}),
we find that $\dot y x^2$ asymptotically tends to a constant.
Substituting this result into Eq.~(\ref{Bdef}) for $B$, we see that
$\dot x$ must tend to a constant.  It is then straightforward to
obtain the asymptotic solution
\begin{eqnarray}
    x &=&  {2 \kappa v_0 \over y_0} (t - t_0) + O(t^{-1}) \cr
    y &=&  y_0 - {\beta \over t} + O(t^{-2})  
\label{asymsolution}
\end{eqnarray}
where $y_0$, $t_0$, $v_0$, and $\beta$ are constants of integration.
Of these constants, $t_0$ simply corresponds to a shift of the zero of
time, while $\beta$ is related to the others by
\begin{equation}
       \beta = {y_0 (a^2 +y_0^2) \over 4 \kappa v_0} \, {B \over E}
 \, \,.
\end{equation}

The meanings of $v_0$ and $y_0$ are clarified by writing the
asymptotic solution in terms of $b$ and $r$:
\begin{eqnarray}
    r &=& v_0 |t| + \dots   \cr \cr
    b &=& {\kappa v_0^2 \over y_0^2} t^2 + \dots    \, \,.
\label{asymRBsolution}
\end{eqnarray}
We see that asymptotically the massive monopoles move with a constant
relative velocity $v_0$, while $b$ varies quadratically with
time.\footnote{ This would imply that as $t$ approaches $\infty$
($-\infty$), the cloud parameter increases (decreases) at a rate that
exceeds the speed of light; we will return to this point below.}
Substituting this asymptotic solution back into the energy, we find 
that 
\begin{equation}
    E = {\mu \over 2} \dot r^2 + {\kappa\over 2}\, {\dot b^2 \over b}
              + \cdots    \, \,.
\end{equation}
As $t \rightarrow \pm \infty$ the terms indicated by dots vanish,
while the two terms shown explicitly tend to constants.  These latter
two terms can be interpreted as a massive monopole energy
$E_r$ and a cloud energy $E_b$, with 
\begin{equation}
     y_0^2 = a^2 {E_r \over E_b}  \, \,.
\end{equation}

Because of the multiple mapping of $b$ and $r$ into the $x$-$y$ plane,
for each asymptotic solution in Eq.~(\ref{asymsolution}) there are seven
other physically equivalent solutions that are obtained by reversing the
signs of $x$ or $y$ or by the interchange of $x$ and $y$.

\subsection{Joining the asymptotic solutions when ${\bf J}=Q=0$ }

Having found the possible asymptotic behaviors, we now want to obtain
solutions of Eq.~(\ref{newELeqs}) for finite times.  As with the
asymptotic solutions, these will depend on four constants of
integration.  One of these can be absorbed in a rescaling of $t$ (with
a corresponding rescaling of $E$ and $B$) and a second in a shift of
the zero of time.  Hence, the trajectories in the $x$-$y$ plane depend
nontrivially only on two constants, which we take to be $B/E$ and
$E_r/E_b$; without loss of generality, we can choose the trajectories
to begin in the first octant, so that the latter quantity is equal to
$y_0^2/a^2$.  Equations~(\ref{xyenergy}) and (\ref{Bdef}) can be
combined to yield a quadratic equation for $dy/dx = \dot y/\dot x$.
The solution of this equation gives $dy/dx$ as a function of $x$, $y$,
and $B/E$.  Hence, two trajectories with the same value of 
$B/E$ but different $y_0$ cannot cross.  While our numerical solutions
indicate that the same may be true for trajectories with the same
$y_0$ and different $B/E$, we have no proof of this statement.

We used numerical integration to obtain solutions for various values
of $y_0$ and $B/E$.  We will begin by describing our results for $y_0
=a$ (i.e., initially equal values of $E_r$ and $E_b$).  In Fig.~1, we
show some trajectories in the $x$-$y$ plane for $y_0 =a$ and several
values of $B/E$.  Figure~2 shows $r$ and $b$ as functions of time for
these trajectories.  Note that $r$ and $b$ have been plotted in units
of $r_c= 2\kappa/\mu$; recall that this is approximately equal to the
sum of the core radii of the two massive monopoles.

\begin{figure}[hbtp]\begin{center}
\PSbox{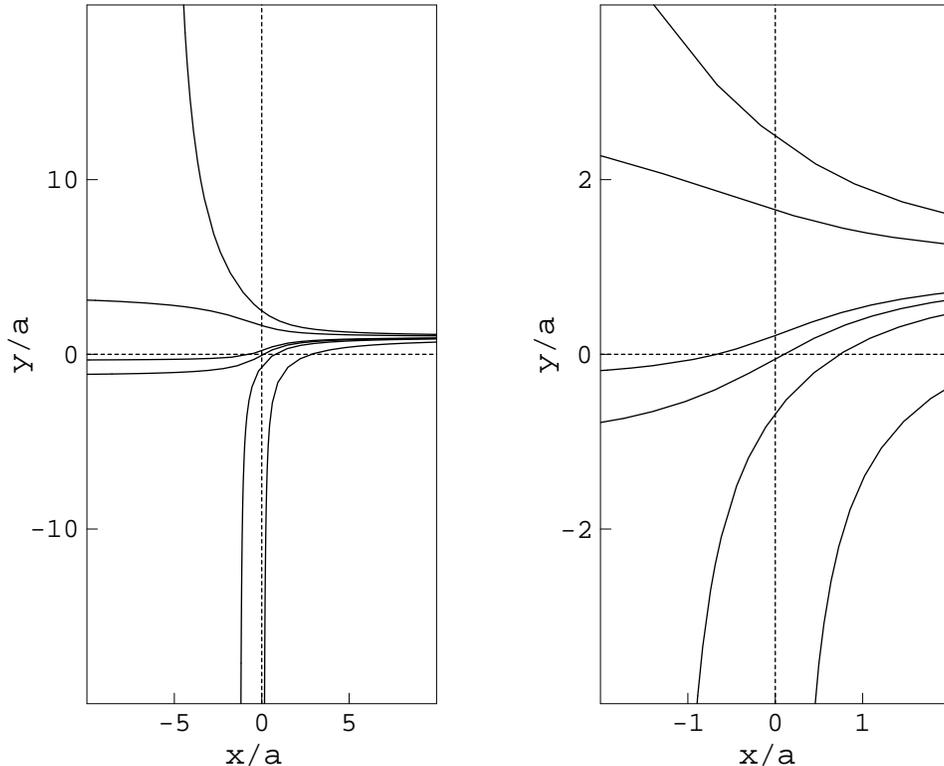
hscale=80 vscale=80  hoffset=-180    
voffset=-35}{3.5truein}{3.5truein}
\end{center} 
\medskip
\caption{
Several trajectories in the $x$-$y$ plane for $y_0=a$.  Beginning from
the top, these correspond to values of $B/E$ of -1.40, -0.60, 0.66,
0.85, 1.20, and 3.00.  The graph on the right is a blow-up of the
region near the origin. 
}
\label{fig:xyplot}
\end{figure}

\begin{figure}[hptb]\begin{center}\PSbox{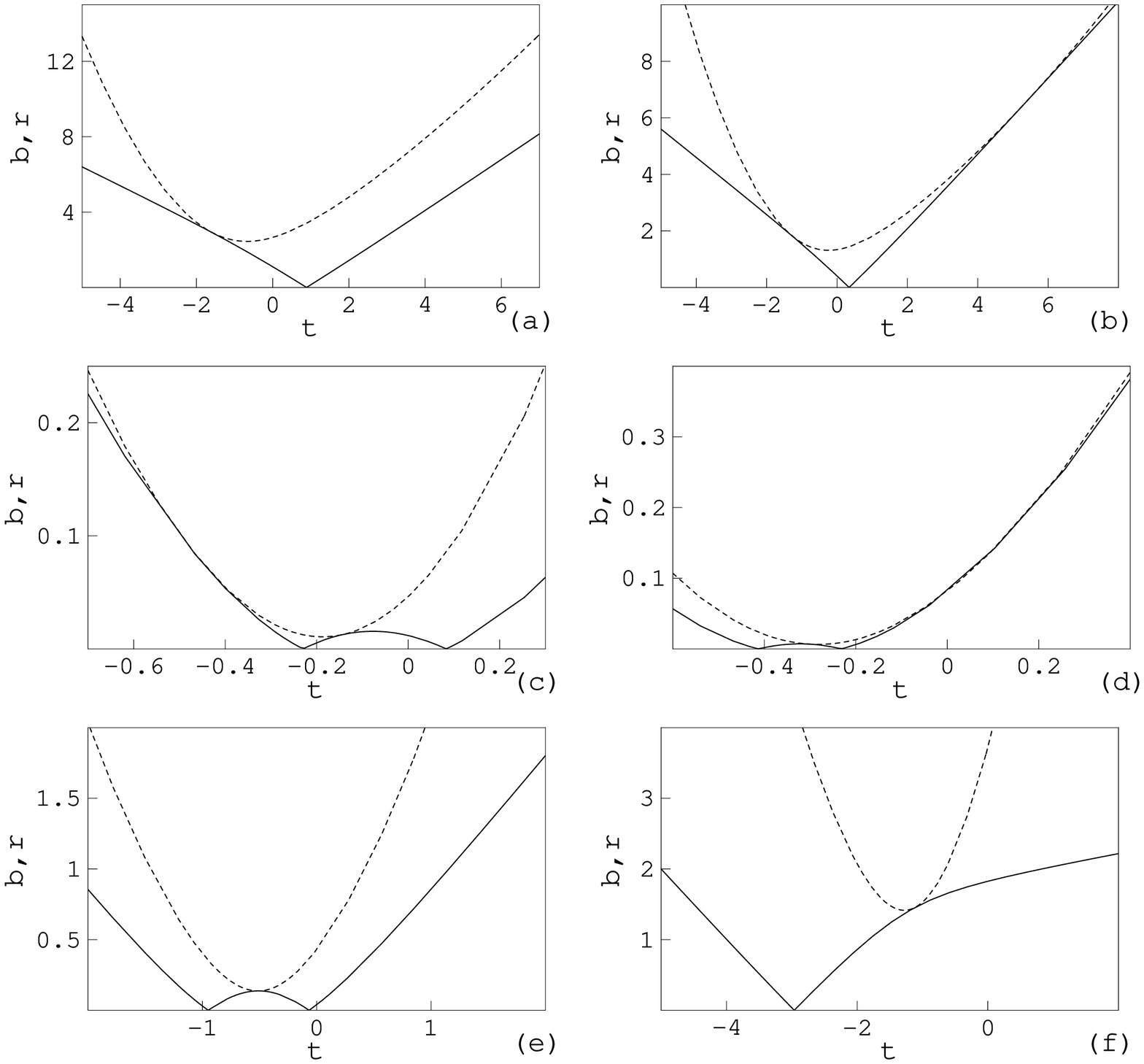
hscale=80 vscale=80 hoffset=-90 voffset=-35}{6truein}{5truein}\end{center}
\medskip
\caption{ Plots of $b$ (dashed line) and $r$ (solid line) as functions
of $t$ for the trajectories shown in Fig.~1.  The progression from (a)
to (f) corresponds to moving from top to bottom in Fig.~1.  Time is
shown in units of $a/2\sqrt{E}$, while $b$ and $r$ are plotted in
units of $r_c=2\kappa/\mu$.  Note that the scale varies from plot to
plot.  }
\label{fig:rbplots}
\end{figure}

In examining these solutions, it is useful to focus on the points
where $b=r$ (corresponding to crossing one of the 45-degree lines
$x=\pm y$) and where $r=0$ (corresponding to crossing either of the
coordinate axes in the $x$-$y$ plane).
When $b=r$, the cloud is of minimal size,\footnote{This is
not necessarily the minimum value of $b$.  In most cases $r<b$ when
$b$ reaches its minimum, so that the smallest value of $b$ corresponds
to a nonminimal cloud.}
and can be viewed as having
shrunk to the line segment joining the massive monopoles.  The field
configuration is an embedding of a purely Abelian ${\rm
SU}(3)  \rightarrow {\rm U(1)}\times {\rm U(1)}$ solution, and
the component of the field in the unbroken SU($N$) is a pure dipole
field.  In terms of massless monopole positions, this corresponds to
a configuration where all of the massless monopoles are located on top
of one of the massive monopoles.  
All trajectories pass through at least one such minimal cloud
configuration.  If $|B/E| < 1$, the trajectory passes through two
such configurations.  Whether $b=r$ once or twice does not appear to have
a
dramatic influence on the large time behavior of the solutions.

The points on a trajectory where $r=0$ are, of course, the points
where the two massive monopoles pass through one another.  Since $x$
and $y$ both vary monotonically, it is clear that every trajectory
must cross at least one coordinate axis and so must have at least one
such point.  Thus, the two massive monopoles that are approaching each
other on a straight line (because ${\bf J}=0$) cannot stop and reverse
direction before they meet.  However, it is possible for them, after
passing through each other once, to stop, reverse direction, pass
through each other a second time, and finally emerge with their
original direction of motion reversed.  Whether this second
possibility is realized is determined, for a given $y_0$, by $B/E$.
For $y_0 = a$, the monopoles reverse direction if $0.51 \lesssim B/E
\lesssim 2.60$; the boundaries of this range correspond to solutions
where the massive monopole velocities vanish as $t \rightarrow
\infty$.

The trajectories displayed in Figs.~1 and 2 were chosen to 
illustrate the various possible behaviors ($b=r$ once or twice, $r=0$
once or twice).   
Several points should be noted.  First, there does not appear to be
any particular striking
effect when $r=0$.  Because $r$ is by definition positive,
$\dot r$ changes sign, but the magnitude of $\dot r$ varies
continuously.  Similarly, $b(t)$ shows no unusual behavior when $r$
vanishes.  All this is consistent with our expectations; since the two
massive monopoles have no direct interactions, they should be able to
pass through each other with little effect. 

By contrast, each of the massive monopoles does interact directly with
one of the massless monopoles comprising the cloud.  This might
suggest that there should be a noticeable effect when a massive and a
massless monopole position coincide; i.e., when $b=r$.  On the other
hand, the massless monopoles are hardly point particles, since the
entire cloud region can be viewed as being composed of the massless
cores.  Nevertheless, it does seem that the greatest effect on the
trajectories occurs when $b \approx r$.  Thus, the largest changes in
the slope of $r(t)$ coincide almost exactly with the vanishing of
$b-r$.  Note also that the effect of these massive-massless
``collisions'' seems to be greatest when they occur at small $b$ and
$r$.  This is perhaps clearest in the trajectories where the
interaction is strong enough to reverse the direction of motion of the
massive monopoles, so that $r=0$ twice.  In these cases, $b$ and $r$
coincide when $r$ is so small that the two massive cores overlap each
other.

\begin{figure} \begin{center}
\PSbox{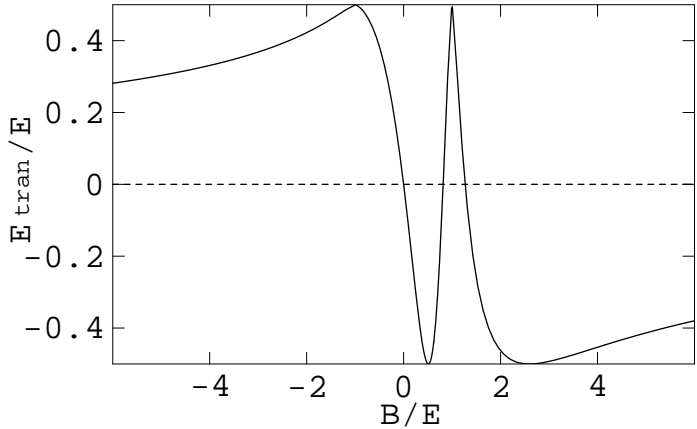
hscale=100 vscale=100 hoffset=-10
voffset=-15}{6truein}{2truein}\end{center}
\medskip
\caption{Energy transfer from the cloud to the massive monopoles as a
function of $B/E$, for $y_0=a$. }
\label{fig:deltaE}
\end{figure}

Even when the massive monopoles do not change direction, we can gauge
the strength of their interaction with the cloud by the amount of
energy transferred  between the cloud and the
massive monopoles.  In Fig.~3 we show this as a function of $B/E$ for
$y_0=a$.  Although the details of this plot vary somewhat with $y_0$, 
two points are fixed:

1) If $B/E=0$, $\dot y$ vanishes identically, so $y$ is constant in
   time and there is no energy transfer.

2) If $B/E = \pm 1$, the trajectory asymptotically approaches one of
   the lines $x=\pm y$ at large time, corresponding to a final state
   with minimal cloud, $b=r$.  In this case, all of the initial cloud
   energy is transferred to the massive monopole kinetic energy.

Figure~3 should be compared with Fig.~4, where we we plot the minimum
value of $b$ as a function of $B/E$.  We see that the largest energy
transfers are associated with small values of $b$, and that the energy
transfer decreases as $b_{\rm min}$ increases for $|B/E|\gg 1$.  In
Fig.~4 we have also shown the value of $b-r$ at the time that $b$
achieves its minimum.  Note that this value is always less than or
comparable to the monopole core radius.

\begin{figure} \begin{center}\PSbox{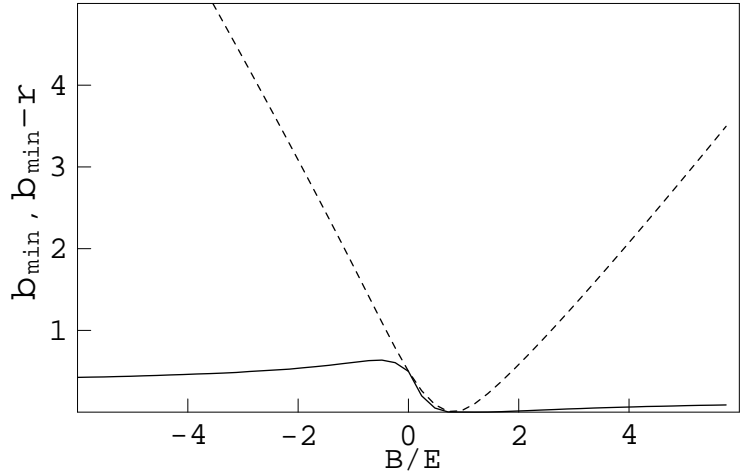
hscale=100 vscale=100 hoffset=-10
voffset=-15}{6truein}{2truein}\end{center}
\medskip
\caption{Minimum cloud parameter $b_{\rm min}$ (dashed line) and the
corresponding value of $b_{\rm min}-r$ (solid line) as functions of
$B/E$, for $y_0=a$.  Both $b$ and $r$ are plotted in units of
$2\kappa/\mu$.}
\label{fig:bmin}
\end{figure}

We have also studied solutions with $y_0 = 100a$ (i.e., $E_r \gg E_b$)
and $y_0 =0.01a$ ($E_r \ll E_b$).  The general picture is very much as
for the case of $y_0=a$.  The analogues of Figs.~1 and 2 are
qualitatively quite similar.  We again find that $b-r$ is small when
the cloud parameter $b$ achieves its minimum size, and the greatest
amount of energy transfer between the cloud and the massive monopoles
occurs when $b_{\rm min}$ is small.  One notable difference is found
in those $y_0=0.01a$ solutions where the massive monopoles reverse
direction.  In contrast with the $y_0=a$ case, the massive monopoles
need not be overlapping at the time of this reversal.  This is readily
understood.  Because the cloud energy is so much greater than the
massive monopole energy, relatively little energy transfer is required
to reverse the sign of $\dot r$, and so $b$ at the time of the
collision can be larger than it was in the previous case.  This
emphasizes that the critical factor for the $y_0=a$ reversal was that
$b$ was small; the fact that the massive cores overlapped was
coincidental.  

\subsection{Solutions with nonzero ${\bf J}$ and $Q$ }

\begin{figure} \begin{center}\PSbox{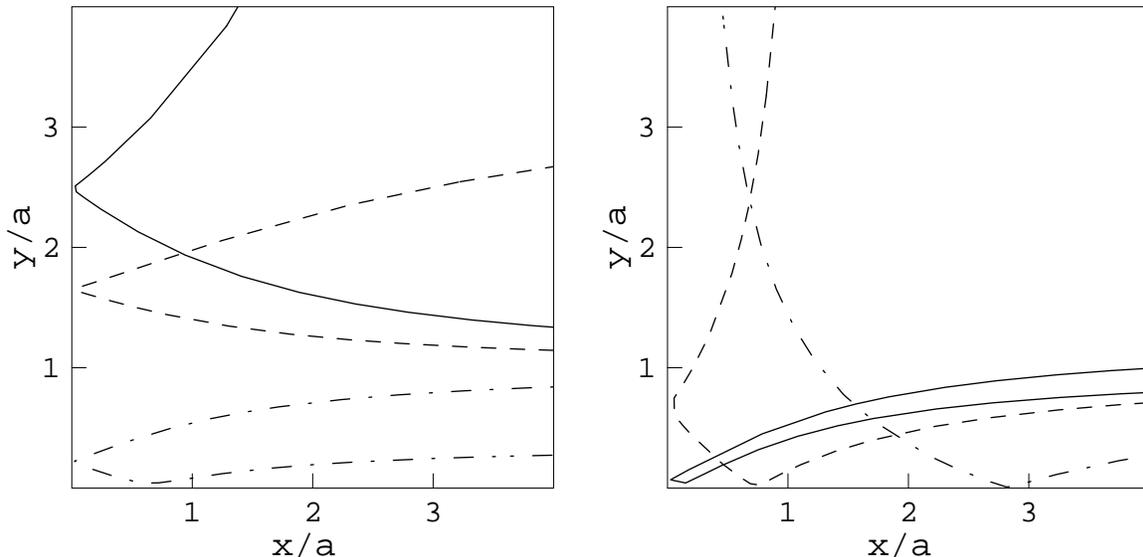
hscale=90 vscale=90 hoffset=-90 voffset=-35}{6truein}{2truein}\end{center}
\bigskip
\caption{Several trajectories in the $x$-$y$ plane for nonzero $\bf
J$.  The values of $y_0$ and $B/E$ are the same as in Fig.~1, while
$J=0.001 a\sqrt{E}$.  The trajectories in the left box correspond
to the top three trajectories in Fig.~1, while those in the right box
correspond to the bottom three trajectories in Fig.~1.  }
\label{fig:jne0}
\end{figure}

The effective potential terms in Eq.~(\ref{Leffective}) come into play
when either $\bf J$ or $Q$ is nonzero.  With nonvanishing angular
momentum, there is a centrifugal barrier that prevents the vanishing
of $r$. The trajectory in the $x$-$y$ plane cannot cross either axis,
and so is confined to a single quadrant.  For large $\bf J$, this
barrier ensures that $r$, and hence $b$, is always large, and thus
suppresses interactions between the massive monopoles and the cloud.
For small but nonzero $\bf J$, the effect of the centrifugal barrier
is to make the trajectories appear to reflect off the $x$- and
$y$-axes.  We illustrate this in Fig.~5.  Except for the value of $\bf
J$, the initial data for the trajectories in this figure are the same
as for those shown in Fig.~1.  Equation~(\ref{Jsquare}) for ${\bf
J}^2$ implies that the motion of the massive monopoles is confined to
a plane.  In Fig.~6, we illustrate this motion by showing the path of
$\bf r$ in the plane perpendicular to $\bf J$ for one of the
trajectories shown in Fig.~5.  Note once again that there is little
evidence of interaction when the massive monopoles are at their
closest approach.

\begin{figure} \begin{center}\PSbox{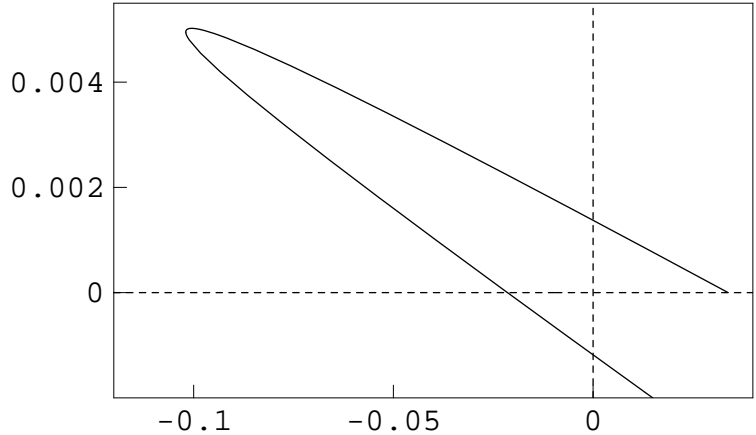
hscale=100 vscale=100 hoffset=-40
voffset=-25}{6truein}{2truein}\end{center}
\medskip
\caption{A portion of the orbit of the massive monopole separation
$\bf r$ in the plane perpendicular to $\bf J$.  Note that the
turnaround of the monopole motion does not occur at the time of
closest approach.  The trajectory shown here corresponds to 
$B/E=1.2$, $y_0=a$, and $J=0.002 a\sqrt{E} $.   }
\label{fig:realspace}
\end{figure}

If $Q \ne 0$, there is a barrier that forbids the vanishing of $b-r$,
or equivalently, the crossing of the lines $x = \pm y$.  When $Q$ is
large, $b-r$ can never be small, and so there is little interaction
among the monopoles.  For small nonzero $Q$, we find behavior analogous to
that for ${\bf J} \ne 0$, with trajectories appearing to bounce off
the lines $x = \pm y$. 

\section{Concluding remarks}

In this paper we have studied the properties of the magnetically
charged counterparts of the electrically-charged massless particles
that arise when a gauge theory is spontaneously broken to a
non-Abelian subgroup.  Previous studies of static BPS solutions have
shown that these massless monopoles coalesce into a cloud of
non-Abelian field surrounding one or more massive monopoles.  In the
scattering solutions that we have studied, these clouds act very much
like thin shells, with a size of order $b$, that contract until they
collide with and bounce off the cores of the massive monopoles.  At
large times, when this shell is far from the massive cores, the cloud
and the massive monopoles are essentially decoupled, and have
separately conserved energies.  Their interaction occurs primarily at
the time that the shell overlaps the massive cores.  We find that the
strength of this interaction, as measured, e.g., by the amount of
energy transfer, tends to be greatest if $b$ is small at the
time of this overlap.

Further evidence in support of this picture can be obtained from
another system that has been studied in some detail.  When SU(3) gauge
theory is spontaneously broken to SU(2)$\times$U(1) there are two
species of fundamental monopoles, one massive and one massless.  The
Nahm data and moduli space metric for the BPS solutions containing one
massless and two massive monopoles were obtained by
Dancer \cite{dancer1,dancer2}.  These 
solutions depend on twelve parameters.  Ten of these correspond to the
center-of-mass position, global SU(2)$\times$U(1) phases, and overall
spatial rotations.  The remaining two, $k$ and $D$, arise as
parameters in elliptic functions that enter the Nahm data.  These span
a geodesic submanifold which, after a nonlinear change of variables to
eliminate coordinate singularities in the moduli space metric, is
often illustrated as in Fig.~7.  (This figure is actually a six-fold
covering of the submanifold, completely analogous to our eight-fold
mapping of $b$ and $r$ onto the $x$-$y$ plane.)

\begin{figure} \begin{center}\PSbox{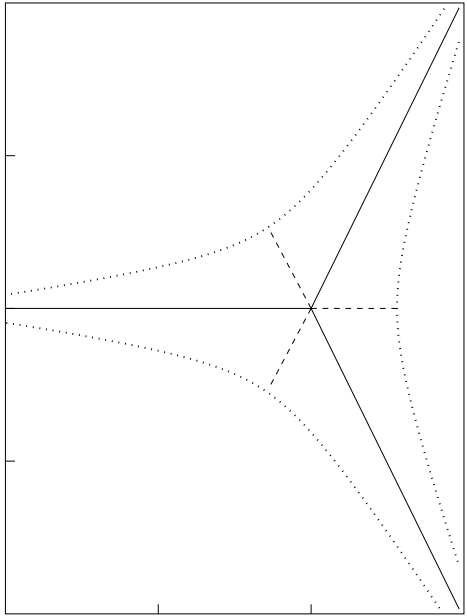
hscale=40 vscale=40 hoffset=125 voffset=0}{6truein}{3truein}\end{center}
\medskip
\caption{The two-dimensional Dancer submanifold discussed in the text.
The hyperbolic solutions correspond to points on the 
solid lines, while the trigonometric solutions lie on the dashed
lines.  The boundaries, indicated by dotted lines, correspond to
embedded SU(2) two-monopole solutions.}
\label{fig:dancer}
\end{figure}

The precise relations between position in this figure and cloud size
and monopole separation are rather complex, with analytic
expressions only known asymptotically \cite{irwin}.  However, the
qualitative picture is easily described.  Points far out along the
three legs of the diagram correspond to solutions containing two
well-separated massive monopoles, with the separation increasing with
distance along the leg, while points in the central part of the figure
correspond to solutions in which the massive monopole cores overlap
and deform each other. The boundaries of the figure correspond to
embedded SU(2) two-monopole solutions; in our language, these are
solutions with infinite cloud parameter.  (These boundaries are
infinitely far, in metric distance, from any interior point.)  The
lines that bisect the legs correspond to special axially symmetric
solutions \cite{dancer2}.  The solid portions of these lines in Fig.~7
correspond to ``hyperbolic
solutions'' that can be viewed as solutions with minimal cloud (the
analogue of our $b=r$ solutions) and with the separation between the
massive monopoles increasing with distance from the center.  The
dashed portions of these lines correspond to ``trigonometric
solutions'' that are composed of two overlapping massive monopoles and
a massless cloud that increases from minimal size at the center to
infinite size at the boundary of the figure.

Geodesics on this two-dimensional submanifold correspond to scattering
solutions with vanishing angular momentum and internal charges.  In
contrast to the SU($N$) example we have studied in this paper, there
are direct interactions between the two massive monopoles.
Consequently, there is nontrivial scattering even when the cloud is
infinitely large.  Thus, each of the three geodesics that bound the
moduli space in Fig.~7 corresponds to a process in which the two
massive monopoles start infinitely far apart, approach head-on, and
then recede at right angles to their initial motion; during the entire
process the cloud remains infinitely large.

Numerical solutions of the geodesic equations were obtained by Dancer
and Leese \cite{dancerleese}.  Examining these, we see behavior quite
consistent with our interpretation of our SU($N$) solutions, in that
the interaction between the cloud and the massive monopoles is greater
if the massive monopoles are closer together at the time of the
minimal cloud configuration.  We see this in two ways.  First, there
are some geodesics that start in one leg of the figure, cross the line
of trigonometric solutions and begin to move down another leg, and
then, after crossing the line of hyperbolic solutions, cross a second
line of trigonometric solutions and exit via third leg.  These may be
viewed as analogous to the SU($N$) scattering solutions in which the
two massive monopoles reverse direction.  As with the latter
solutions, these ``three-leg trajectories'' are found only if the
massive monopoles are sufficiently close in the minimal cloud
configuration; in other words, if the trajectory goes too far down the
second leg, the massive monopoles cannot reverse direction.
 
The effects of cloud interactions on the ``two-leg trajectories'' that
cross a line of hyperbolic solutions but only a single line of
trigonometric solutions are more subtle.  At large times these
trajectories lie between the boundary and the line of hyperbolic
solutions, with the cloud size increasing as the relative distance to
the boundary decreases.  Comparing the distance of the geodesics from
the boundary at points equally far down the initial and final legs
thus gives a measure of the net interaction between the cloud and the
massless monopoles, somewhat analogous to the SU($N$) energy transfer
that we plotted in Fig.~3.  Examination of the Dancer-Leese
geodesics shows that the greatest change between initial and final
legs, and thus the greatest interaction, occurs if the geodesic
crosses the line of hyperbolic solutions nearest the center of the
figure; i.e., nearest the point of minimal massless monopole
separation. 

Before closing, we must address the validity of the moduli
space approximation.  There is a potential problem in this regard in
any theory where the excitation of massless particle modes is a
possibility.  The fact that the asymptotic solution of
Eq.~(\ref{asymRBsolution})
has the cloud parameter $b$ increasing faster than the speed of light
at large times is an indication that there actually is a breakdown of
the moduli space approximation in the example we have studied.
Considerable insight in this issue can be obtained by studying
\cite{MSApaper} an SO(5) example with a spherically symmetric cloud.
This symmetry makes it possible to compare the predictions of the
moduli space approximation with a numerical solution of the full field
equations, starting from a slowly varying initial configuration.  It
turns out that the moduli space approximation continues to be reliable
until $\dot b$ becomes of order unity.  After that time, the edge of
the cloud develops into a rather well-defined front expanding outward
at the speed of light.  Behind this front, in the cloud interior, the
time evolution of the fields continues to be well-described by the
moduli space solution (i.e., as if $b \sim t^2$).  Ahead of the front,
the evolution of the fields is much slower, consistent with a slow linear
growth of the cloud at a rate fixed by the initial value of $\dot b$. 

How does this affect the picture of monopole-cloud scattering that we
have developed?  Two cases can be distinguished.  In one, a static
solution with finite $r$ and $b$ is subject to an external disturbance
that induces small, but nonzero, values for $\dot r$ and $\dot b$.  As
long as $\dot b$ remains small (energy conservation ensures that $\dot
r$ will remain small) the moduli space description of the interaction
between the cloud and the massive monopoles should be reliable.
Eventually $\dot b$ will become of order unity, but since $r \ll b$
when this happens, the moduli space approximation should continue to
give a good description of the subsequent motion of the massive
particles.  Thus the main modification to our picture is that the
velocity of the cloud asymptotically approaches unity, rather than
increasing without bound.

In the other case, the initial low-energy configuration approximates a
solution with large cloud that is contracting with the magnitude of
$\dot b$ being of order unity. We expect that the
deviation from the moduli space solution will lead to the emission of
long wavelength, low amplitude waves of the massless boson fields that
propagate at the speed of light.  This will leave behind a
configuration whose evolution then proceeds much as in the previous
case. 

In closing, we note that while our focus in this paper has been on the
massless monopole clouds in classical multimonopole configurations,
much of the underlying motivation was to gain insight into the states
of the electrically-charged elementary particle sector that are dual
to these.  Thus, it would be of great interest to study the BPS and
near-BPS states containing massless and massive elementary particles
with non-Abelian charge with the aim of identifying the counterparts
of the cloud parameter $b$ or of the asymptotic separation of the
energy between the cloud and the massless monopoles.

\acknowledgments 
We thank Kimyeong Lee, Changhai Lu, and Piljin Yi for
helpful discussions.  This work was supported in part by the
U.S. Department of Energy.

\appendix
\section{}

In this appendix we show explicitly how to express the metric
${\cal G}_{\rm rel}$ of Eq.(\ref{zmetric}) in terms of the Euler
angles.  We restrict ourselves here to the case of 
${\rm SU(4)}
\rightarrow {\rm U(1)}\times{\rm SU(2)}\times{\rm U(1)}$. 

A natural set of parameters is the cloud size $b$; the massive
monopole separation $r$; three spatial Euler angles $\theta$, $\phi$,
and $\psi$; three internal SU(2) Euler angles $\alpha$, $\beta$, and
$\gamma$; and the relative U(1) phase $\chi$.  This is a total of nine
but, as we discussed in Sec.~2, we only expect eight parameters.  We
will see later that this conflict is resolved by the identification of a
spatial Euler angle with one of the internal SU(2) Euler angles.

The $z^a_A$ transform as doublets under both the spatial and the internal
SU(2), with the upper and lower indices being spatial and internal,
respectively.  Hence the $z^a_A$ for arbitrary values of the spatial
Euler angles must be of the form 
\begin{eqnarray}
\left( \begin{array}{c} z^1_A\\ z^2_A \end{array} \right) =
U(\theta,\phi,\psi) \left( \begin{array}{c} a_A\\ b_A \end{array} \right)
\end{eqnarray}
where $a_A$ and $b_A$ are functions of $b$, $r$, $\chi$, and the
internal Euler angles.  We choose a parameterization of
$U(\theta,\phi,\psi)$ such that
\begin{eqnarray}
z^a_A=a_A v^a + b_A \tilde{v}^a
\end{eqnarray}
where 
\begin{eqnarray}
v^a&=&\left( \begin{array}{c} \cos \frac{\theta}{2}
e^{\frac{i}{2}(-\psi-\phi)} \\ \sin \frac{\theta}{2}
e^{\frac{i}{2}(-\psi+\phi)} \end{array} \right) \nonumber\\
\tilde{v}^a&=&\left( \begin{array}{c} \sin \frac{\theta}{2}
e^{\frac{i}{2}(\psi-\phi)} \\ -\cos \frac{\theta}{2}
e^{\frac{i}{2}(\psi+\phi)} \end{array} \right)
\end{eqnarray}
Similarly, $a_A$ and $b_A$ can be written as linear combinations of
\begin{eqnarray}
u_A&=&\left( \begin{array}{c} \sin \frac{\alpha}{2}
e^{\frac{i}{2}(-\gamma+\beta)} \\ \cos \frac{\alpha}{2}
e^{\frac{i}{2}(-\gamma-\beta)} \end{array} \right) \nonumber\\
\tilde{u}_A&=&\left( \begin{array}{c} \cos \frac{\alpha}{2}
e^{\frac{i}{2}(\gamma+\beta)} \\ -\sin \frac{\alpha}{2}
e^{\frac{i}{2}(\gamma-\beta)} \end{array} \right)
\end{eqnarray}
with coefficients that depend only on $r$, $b$, and $\chi$.  By
performing a gauge rotation, we can make $a_A$ proportional to $u_A$.
We then have
\begin{eqnarray}
z^a_A=A u_A v^a + B\tilde{u}_A \tilde{v}_A + C u_A \tilde{v}_A
\end{eqnarray}

The 2 relative phases among $A, B$ and $C$ can be absorbed in
redefinitions of $\gamma$ and $\psi$. Then, by a redefinition
$A\rightarrow Ae^{\frac{i}{2} \chi}$, $B\rightarrow Be^{\frac{i}{2}
\chi}$ and $C\rightarrow Ce^{\frac{i}{2} \chi}$, we can make the
$A,B,C$ all real and non-negative.  Equations~(\ref{rFROMz}) and
(\ref{bFROMz}) then give the constraints
\begin{eqnarray}
A^2+C^2 &=& \frac{1}{2}(b+r) \nonumber\\
B^2 &=& \frac{1}{2}(b-r)\nonumber\\
BC &=& 0
\end{eqnarray}
These are solved by $A=\frac{1}{2} \sqrt{b+r}$, $B=\frac{1}{2}
\sqrt{b-r}$, $C=0$.  This gives us 
\begin{eqnarray}
    z^1_1 &=& \left(\sqrt{{b+r} \over 2} \sin {\alpha \over 2} \cos 
    {\theta \over 2} e^{-{i \over 2}(\gamma+\psi)} + 
              \sqrt{{b-r} \over 2} \cos {\alpha \over 2} \sin 
    {\theta \over 2} e^{{i \over 2}(\gamma+\psi)}\right) 
    e^{{i \over 2}(\beta +\chi -\phi)},
    \nonumber\\
    z^1_2 &=& \left(\sqrt{{b+r} \over 2} \cos {\alpha \over 2} \cos 
    {\theta \over 2} e^{-{i \over 2}(\gamma+\psi)} - 
              \sqrt{{b-r} \over 2} \sin {\alpha \over 2} \sin 
    {\theta \over 2} e^{{i \over 2}(\gamma+\psi)}\right)
    e^{{i \over 2}(-\beta +\chi -\phi)},
    \nonumber\\
    z^2_1 &=& \left(\sqrt{{b+r} \over 2} \sin {\alpha \over 2} \sin 
    {\theta \over 2} e^{-{i \over 2}(\gamma+\psi)} - 
              \sqrt{{b-r} \over 2} \cos {\alpha \over 2} \cos 
    {\theta \over 2} e^{{i \over 2}(\gamma+\psi)}\right)
    e^{{i \over 2}(\beta +\chi +\phi)},
    \nonumber\\
    z^2_2 &=& \left(\sqrt{{b+r} \over 2} \cos {\alpha \over 2} \sin 
    {\theta \over 2} e^{-{i \over 2}(\gamma+\psi)} + 
              \sqrt{{b-r} \over 2} \sin {\alpha \over 2} \cos 
    {\theta \over 2} e^{{i \over 2}(\gamma+\psi)}\right)
    e^{{i \over 2}(-\beta +\chi +\phi)}.
\label{relations}
\end{eqnarray}
Notice that $\gamma$ and $\psi$ always come together in the combination
$\gamma+\psi$. This is a manifestation of the axial symmetry of
this configuration: for any spatial axial
rotation that changes the parameter $\psi$, one can do
an opposite axial gauge rotation that changes $\gamma$, so that the
metric is invariant.

Substituting these relations into the Eq.~(\ref{zmetric}) 
gives
\begin{eqnarray}
    {\cal G}_{\rm rel} &=& \left(\frac{1}{2} \mu + \frac{\kappa}{2}
    \frac{b}{b^2-r^2} \right) dr^2 + \frac{\kappa}{2}
    \frac{b}{b^2-r^2} db^2 - \kappa \frac{r}{b^2-r^2} db dr
    \nonumber\\
       &+&\frac{1}{2}(\mu r^2+\kappa
    b)(\sigma_1^2+\sigma_2^2)+\frac{1}{2}\kappa
    b(\tau_1^2+\tau_2^2)+\kappa
    \sqrt{b^2-r^2}(-\sigma_1\tau_1+\sigma_2\tau_2) 
    \nonumber\\
       &+&\frac{\kappa}{2}\left( b-\frac{\mu r^2}{\kappa+\mu
    b} \right) (\sigma_3+\tau_3)^2 - \frac{\kappa^2 r}{\kappa+\mu
    b}(\sigma_3+\tau_3)d\chi + \frac{\kappa^2 b}{2(\kappa+\mu b)}
    d\chi^2
\label{explicitmetric}
\end{eqnarray}
where
\begin{eqnarray}
\sigma_1 &=& -\sin \psi \, d\theta + \cos \psi \sin \theta \, d\phi,
    \nonumber\\
    \sigma_2 &=& \cos \psi  \, d\theta + \sin \psi \sin \theta \, d\phi,
    \nonumber\\
    \sigma_3 &=& d\psi + \cos \theta \, d\phi,
\label{su2sigmas}
\end{eqnarray}
and 
\begin{eqnarray}
\tau_1 &=& -\sin \gamma  \,d\alpha + \cos \gamma \sin \alpha  \,d\beta,
    \nonumber\\
    \tau_2 &=& \cos \gamma  \,d\alpha + \sin \gamma \sin \alpha  \,d\beta,
    \nonumber\\
    \tau_3 &=& d\gamma + \cos \alpha \,d\beta,
\label{su2taus}
\end{eqnarray}
are the left-invariant one-forms for the rotational and internal SU(2)'s,
respectively.    

There are several interesting limits of this metric.  If the relative
distance between the two massive monopoles ${\bf r}$ and the relative
U(1) charge $Q$ both vanish, it reduces to the metric for one massive
and one massless monopole in SO(5) \cite{so5,kwy}:
\begin{eqnarray}
{\cal G}_{\rm SO(5)} = \frac{\kappa}{2}\left[ \frac{db^2}{b} +
b(\tau_1^2+\tau_2^2+\tau_3^2) \right] .
\label{so5metric}
\end{eqnarray}

In the limit where we have minimal cloud $b=r$ and vanishing internal
SU(2) charges, we get the metric for two distinct
massive SU(3) monopoles \cite{twomono}.  The angles $\psi$ and
$\gamma$ can be absorbed into a redefinition of $\chi$, and we obtain
\begin{eqnarray}
{\cal G}_{\rm SU(3)} = \frac{1}{2} \left( \mu+\frac{\kappa}{r} \right)
\left[ dr^2+r^2(\sigma_1^2+\sigma_2^2) \right] + \frac{\kappa^2}{2}
\frac{r}{\kappa+\mu r} (-d\chi+\cos\theta\, d\phi)^2 \, .
\end{eqnarray}

We can also consider the limit where $\mu$ is much greater than 
both $\kappa/b$ and the initial energy of the
cloud.  To leading order, the massive monopole motion decouples from
that of the cloud, and we may set $\dot{\bf r}=0$.
Making the change of variables $b=\frac{u^2}{4}$ and
$r=\frac{\zeta}{2}$, we get the Eguchi-Hanson \cite{Eguchi}
metric\footnote{We thank Kimyeong Lee for pointing this out to us.}.
\begin{eqnarray}
{\cal G}_{\rm EH} = \frac{\kappa}{2} \left
\{ \frac{du^2}{1-\frac{4\zeta^2}{u^4}} + \frac{u^2}{4}\left[
\tau_1^2+\tau_2^2+ \left(1-\frac{4\zeta^2}{u^4}\right) \tau_3^2 \right]
\right\} \, .
\end{eqnarray}
When $\zeta=0$, this reduces to Eq.~(\ref{so5metric}).

In order to study the classical dynamics for the general case with
nonzero angular momenta and internal charges, it is most convenient
to first eliminate the redundant parameters.  This can be done,
e.g., by simply choosing $\psi=0$.  We then write the
angular part of the metric as
\begin{eqnarray}
{\cal G}_{\rm ang} = I_{ij}(r,b) \omega^i \omega^j,
\label{angularpart}
\end{eqnarray}
where the $\omega_i$ ($i=1, \dots,6$) are one-forms expressed in terms
of the various rotational and gauge angles.  These satisfy an
algebra of the form 
\begin{eqnarray}
d\omega^i=\frac{1}{2}C^i_{jk} \, \omega^j \wedge \omega^k.
\label{algebra}
\end{eqnarray}
where the $C^i_{jk}$ are antisymmetric in $j$ and $k$. 
We then define the charges to be $X_i=I_{ij}\omega^j$. Using
Eq.~(\ref{algebra}), we can write the angular equations of motion for
Eq.~(\ref{angularpart}) as
\begin{eqnarray}
\frac{d}{dt} X_i + C^j_{ki} (I^{-1})^{kl} X_j X_l =0.
\label{JdotEq}
\end{eqnarray}

Ordinarily, we could have chosen the $\omega^j$ to include the three
$\sigma_a$ and the three $\tau_a$.  However, after eliminating the
redundant angular variable, this is not possible.  Instead, we choose
\begin{eqnarray}
\omega^1 &=& -\sin\theta \,d\phi , \qquad 
\omega^2 = d\theta , \qquad
\omega^3 = \tau_1 , \cr 
\omega^4 &=& \tau_2 , \qquad
\omega^5 = \tau_3+\cos\theta \,d\phi ,\qquad
\omega^6 = d\chi \, .
\label{one-forms}
\end{eqnarray}
With this choice, the $C^i_{jk}$ are not constants, as they are in
more familiar examples.  The nonzero $C^i_{jk}$ are given by
\begin{eqnarray}
C^1_{21}&=&C^3_{41}=C^4_{13}=\cot\theta  \cr
C^3_{45}&=&  C^4_{53} = C^5_{21}= C^5_{34}   =1 \, .  
\end{eqnarray}
$I_{ij}$ can be obtained from Eq.~(\ref{explicitmetric}) for the
metric.  It is a block diagonal matrix formed by three two-by-two
blocks, with the first two blocks being identical.  It is
straightforward to verify that Eq.~(\ref{JdotEq}) implies that the square
of
the angular momentum, Eq.~(\ref{Jsquare}), and the square of the internal
SU(2) charge, Eq.~(\ref{Tsquare}), are indeed constant.


\begin{thebibliography}{}

\bibitem{montonen} C. Montonen and D. Olive, Phys. Lett. {\bf 72B}, 117
(1977); H. Osborn, Phys. Lett. {\bf 83B}, 321 (1979).

\bibitem{bps} E.B. Bogomol'nyi, Sov. J. Nucl. Phys. {\bf 24}, 449
(1976); M.K. Prasad and C.M. Sommerfield, Phys. Rev. Lett. {\bf 35},
760 (1975).

\bibitem{Gindex} E.J. Weinberg, Nucl. Phys. {\bf B167}, 500 (1980).

\bibitem{so5} E.J. Weinberg, Phys. Lett. {\bf B119}, 151 (1982).

\bibitem{nonabelindex} E.J. Weinberg, Nucl. Phys. {\bf B203}, 445 (1982).

\bibitem{pathologies} A. Abouelsaood, Phys. Lett. {\bf 125B}, 467 (1983);
P. Nelson, Phys. Rev. Lett. {\bf 50}, 939 (1983);
P. Nelson and A. Manohar,  Phys. Rev. Lett. {\bf 50}, 943 
(1983); A. Balachandran, G. Marmo, M. Mukunda, J. Nilsson,
E. Sudarshan, and F. Zaccaria, Phys. Rev. Lett. {\bf 50}, 1553 (1983);
P. Nelson and S. Coleman, Nucl. Phys. {\bf B237}, 1 (1984).

\bibitem{ewyi} E.J. Weinberg and P. Yi, Phys. Rev. D {\bf 58},
046001 (1998).

\bibitem{msa} N.S. Manton, Phys. Lett. {\bf 110B}, 54 (1982).

\bibitem{kwynonabelian} K. Lee, E.J. Weinberg, and P. Yi, Phys. Rev.   
D {\bf 54}, 6351 (1996).

\bibitem{validity} N.S. Manton and T.M. Samols, Phys. Lett. B {\bf
215}, 559 (1988);  D. Stuart, Commun. Math. Phys. {\bf 166}, 149
(1994). 

\bibitem{twomono} K. Lee, E. Weinberg, and P. Yi, Phys. Lett. B {\bf
376}, 97 (1996); J.P. Gauntlett and D.A. Lowe, Nucl. Phys.  {\bf
B472}, 194 (1996); S.A. Connell, {\it The dynamics of the $SU(3)$
charge (1,1) magnetic monopoles,} University of South Australia
preprint.



\bibitem{kwy} K. Lee, E.J. Weinberg, and P. Yi, Phys. Rev. D {\bf 54},
1633 (1996).

\bibitem{proofs} M.K. Murray, J. Geom. Phys. {\bf 23}, 31 (1997);
G. Chalmers, {\it Multi-monopole moduli spaces for $SU(N)$ gauge
group}, ITP-SB-96-12, hep-th/9605182.

\bibitem{dancer1}  A.S. Dancer, Commun. Math. Phys. {\bf 158}, 545
(1993). 

\bibitem{dancer2}  A.S. Dancer, Nonlinearity {\bf 5}, 1355 (1992). 

\bibitem{irwin} P. Irwin, Phys. Rev. D {\bf 56}, 5200 (1997).

\bibitem{dancerleese} A.S. Dancer and R.A. Leese, Proc. R. Soc. London
{\bf A440}, 421 (1993).

\bibitem{MSApaper} X. Chen, H. Guo, and E.J. Weinberg, in preparation.

\bibitem{Eguchi} T. Eguchi and A.J. Hanson, Ann. Phys. (N.Y.) {\bf
120}, 82 (1979).

\end{thebibliography}
\end{document}